\documentstyle[preprint,eqsecnum,aps,epsf]{revtex}	
\begin{document}

\begin{center}

\Large{\bf Scintillation efficiency of nuclear recoils in a CaF$_{2}$(Eu) crystal
for dark matter search \\}

\large
R. Hazama\footnote{
Present address: Center for Experimental Nuclear Physics and 
Astrophysics, and Department of Physics, University of Washington, Seattle, WA 98195, USA. {\em E-mail:} ryutah@u.washington.edu}, 
S. Ajimura, 
H. Hayakawa, 
K. Matsuoka, 
H. Miyawaki, \\ 
K. Morikubo, 
N. Suzuki, 
T. Kishimoto \\

\normalsize

{\it Department of Physics and Laboratory of Nuclear Studies, \\
Osaka University, Toyonaka, Osaka 560, Japan \\}

\end{center}

\begin{abstract}
The scintillation efficiency (quenching factor) of 
fluorine and calcium nuclei recoiling in a CaF$_{2}$(Eu) crystal 
was measured in an energy region relevant to dark matter searches. 
The recoiling nuclei were obtained via the $^{19}$F(n,n') and 
$^{40}$Ca(n,n') reactions, where the neutron beam was provided 
via the d(d,n)$^3$He reaction. 
The quenching factor of F and Ca nuclei in 
the CaF$_2$(Eu) crystal was found to be 
11 $\sim 20 \%$, and 9 $\sim 23 \%$ relative to the electron equivalent 
energy for 53 $\sim$ 192 keV and 25 $\sim$ 91 keV recoil energies, respectively, 
and the energy dependence was observed. 
The quenching factor we measured here is a little larger than 
that of previous studies, which may depend on the Eu doping concentration 
of the crystal. 
\end{abstract}

\vspace{1cm}
\hspace{-0.5cm}{\it PACS:} 29.40.Mc, 95.35.+d, 14.80.Ly. 

\section{Introduction}
\typeout{SET RUN AUTHOR to \@runauthor}

Weakly Interacting Massive Particles (WIMPs), which are major 
candidates for cold dark matter in the universe (especially in conjuction with proposed 
supersymmetric particles \cite{griest}), 
interact rarely with a nucleus. 
When they do, the nucleus receives a recoil energy 
in the 10 $\sim$ 100 keV range by elastic scattering 
via either the spin-independent or the spin-dependent interaction \cite{witten}. 
Compared to coherent (spin-independent) scattering,  
experiments so far have far below the necessary sensitivity to 
exclude the dark matter candidates that have axial vector coupling with matter 
(spin coupled dark matter) \cite{review}.  
Recently several notable advantages of the spin-dependent interaction channel 
have been noticed \cite{collar}\cite{falk}\cite{klap}. 
Calculations by Ellis et al. \cite{ellis} suggest that fluorine 
provides a particularly high elastic scattering rate for WIMPs that 
have axial coupling. 
The rate estimation for $^{19}$F is fairly stable against the 
calculations of nuclear matrix elements, which in some cases vary 
more than an order of magnitude for other nuclei \cite{jungman}. 
We used fluorine-containing CaF$_2$(Eu) scintillator. 
Use of scintillators allows fabrication 
of multi-kg or heavier detectors. 
Experiments to search for rare events require a 
large quantity of material in a low background environment. 
High statistics is especially 
vital for studying the annual modulation, which is 
a unique signature of halo dark matter in our galaxy. 
We developed a new detector system, ELEGANT VI, which consists 
of 25 CaF$_{2}$(Eu) crystals totaling 7.2 kg. 
Details of ELEGANT VI are given elsewhere \cite{km}\cite{wepa}. 

The light output of scintillators for nuclear recoil is 
known to be lower than that for electrons, 
particularly in the low energy region. 
The observed electron equivalent energy (pulse height) $E_{0}$ is given as
\begin{equation}
E^i_0 = f^i E^i_R,
\end{equation}
where $E_{R}$ is the recoil energy and $f$ is the quenching (conversion) factor of the pulse height for a recoiling nucleus ($i$) with 
respect to the electron. 
It is known that the f-value depends largely on scintillators 
and recoiling nuclei. 
Although the measurements have been carried out by the 
UK(United Kingdom) Collaboration \cite{UK}\cite{UK98} and the BPRS(Beijing/Paris/Rome/Saclay) 
Collaboration \cite{BPRS} for calcium and fluorine,  
these measurements are not consistent with each other. 
The difference may be attributed to the characteristics of each 
scintillator, such as 
the activator concentration, 
the configuration, 
and the PMT photocathode sensitivity. 
Therefore we studied one of the 25 crystals that are used for the dark matter 
experiment. 
This is essential to evaluate our detector's sensitivity to WIMPs. 

\section{Experimental setup}

\subsection{Pulsed-monoenergetic neutrons}

The response of the detector to the nuclear recoil of $^{19}$F and $^{40}$Ca 
was measured, where recoil nuclei were produced 
by the $^{19}$F(n,n') and the $^{40}$Ca(n,n') reactions, respectively. 
The experimental setup is schematically shown in Fig.~\ref{setup}. 
A pulsed 0.75 MeV deuteron beam was provided by the 3.2 MV 
Pelletron Accelerator at the Tokyo Institute of Technology. 
The average beam current was 2 $\mu$A at a repetition rate of 2 MHz.
The $\sim$ 3.7 MeV pulsed neutron beam 
was obtained by the d(d,n)$^3$He reaction. 
The neutron flux was about 2$\times$10$^{7}$ /s/sr. 
The energy spectrum of the neutron beam was measured by a time-of-flight (TOF) 
method between the RF signal (the primary target) and a liquid scintillator 
placed in the forward direction. 
This indicated a spread in neutron energy of around 
0.26 MeV (1 $\sigma$). 
The main contribution to this spread comes from the deuteron energy 
loss in the primary target of Tid$_{2}$. 
The Tid$_{2}$ had a thickness of 0.5 mg/cm$^{2}$ and was made by d$_{2}$ 
gas electrolyzed on a titanium support (which also served as a deuteron beam stop). 
The neutron pulse height response of the BC501A was found to be in agreement 
with the calculation of V. V. Verbinski et al. \cite{verbin}. 
Details of the pulsed neutron beam are described in Ref.~\cite{nagai}. 
Neutrons emitted at 0$^{\circ}$ were collimated using a cylindrical paraffin 
collimator, 60 cm in length with a tapered hole along the beam axis. 
The hole was 1 cm in diameter near the primary target and 3 cm in diameter near 
the secondary target, the CaF$_{2}$(Eu) crystal. 

\subsection{\bf The CaF$_2$ detector}

The CaF$_{2}$(Eu) crystal, a 45 mm cube with a europium 
doping content of 0.17 $\%$ by mass 
was placed at the secondary target point. 
It has a light guide of CaF$_{2}$(pure) crystals coupled to 
photomultipliers \cite{km}\cite{wepa}. 
The CaF$_{2}$(Eu) and CaF$_2$(pure) crystals were made by BICRON. 
The tube is 1.5 inch Hamamatsu H3178 (quartz window).
With this design the CaF$_2$ detector is capable of detecting a low 
energy signal (0.35 keV/photoelectron). 
The whole crystal is covered with four teflon sheets (BICRON BC-642 PTFE 
Reflector tape) and an aluminized mylar 
(Tokyo-metalizing 90313) to maximize the light collection. 
The linearity of the energy calibration for the CaF$_{2}$(Eu) was 
checked using the $^{241}$Am source and light emitting diode(LED). 
Typical energy resolution as measured with the $^{241}$Am 59.5 keV $\gamma$ ray is about $\bigtriangleup E/E$ = 25 $\%$(FWHM). 
The LED (NLPB500 Nichia Chemical) is a blue LED 
and its maximum fluorescence emission-wavelength is close to 
that of the CaF$_2$(Eu). 
This light is sent to both sides of the crystals through optical fibers (BCF98 BICRON). 
The energy calibration and efficiency of signals below 5 keV are difficult to test 
using a radioactive source, so we generated the pulse shape of the low energy signal 
with the LED connected to a shaper amplifier. 
It is verified that the signal has a similar shape to the real gamma ray signal by using a Flash ADC. 
The stability of the system was also monitored periodically by 
lighting the LED with a rate of 10 Hz during the experiment. 

\subsection{\bf The neutron detector and NaI detector}

The signal from the CaF$_2$ detector was measured in coincidence with 
scattered neutrons detected by a BICRON BC501A liquid scintillator, 
20 cm $\phi \times$ 5 cm, viewed by a Hamamatsu R4144 photomultiplier. 
This was placed 2 m from the CaF$_{2}$(Eu) crystal with scattering angles 
of 30$^{\circ}$, 40$^{\circ}$, 50$^{\circ}$, and 60$^{\circ}$. 
Kinematics then defines the recoil energy in the target. 
Corresponding recoil energies are 53 $\sim$ 192 keV and 25 $\sim$ 91 keV for F and Ca, respectively. 
The liquid scintillator is sensitive to both $\gamma$-rays and neutrons. 
They were separated by using a standard pulse shape discrimination 
(PSD) method. 
The typical distribution of the total integrated charge over 100 ns (which depends on 
scintillation decay time) versus the initial integrated charge over 26 ns 
(which is proportional to the rise time) recorded by the ADC is shown in Fig.~\ref{spect}(A). 
The two loci correspond to the $\gamma$ and $n$ events. 
The leakage of the $\gamma$ events to $n$ events in the PSD 
selection is less than 1 $\%$ with a $E_{\gamma}$ ($\gamma$ energy) $>$ 0.7 MeV. 
A heavy metal with a thickness of 10 cm was placed between 
the primary Tid$_{2}$ target and the neutron detector to shield 
neutrons and $\gamma$ rays coming directly from the 
Tid$_{2}$ target. 
Accidental coincidence due to the counting rate of the neutron detector became 
tolerable using this shielding. 
Accidental coincidence background was further reduced by the 
pulse height cut of $\gamma$-rays and neutrons in the neutron detector. 
Only events which produced a recoil proton pulse height in the neutron detector with 
a $E_n$ (neutron energy) $>$ 1.2 MeV were included so as to reduce background coincidences, 
which come from detection of scattered neutrons. 
The neutron detector was calibrated with the $^{88}$Y 898 and 1836 keV 
$\gamma$-ray. 

Two NaI(Tl) scintillators, each with the dimensions of 
$85\times95\times95$ mm$^3$ were placed above and below the 
CaF$_{2}$ scintillator to detect escaped $\gamma$-ray events. 
NaI(Tl) detector was calibrated with $^{22}$Na 511 and 
1275 keV $\gamma$-ray. 
The paraffin collimator prevented the NaI(Tl) detector from 
seeing the primary target position directly. 

\subsection{\bf Electronics}

The decay time of the CaF$_2$ detector is quite long ($\sim3 \mu$sec). 
The output pulses from the PMTs were fed into the high-gain 
Timing Filter Amplifier (TFA), which we developed with an RC circuit 
of integration time-constant 200 ns. 
The signals from the two photomultipliers were separately discriminated 
at the level well below one photoelectron. 
Coincidence of two signals from the left and right PMTs with a 
resolving time of 200 ns made a trigger signal. 
Pulses from two PMTs were summed and fed again into the TFA 
with an integration time-constant 1 $\mu$s. 
This procedure is particularly useful for a low energy signals 
with a few photons scattered during the long decay time. 
Then pulses above a threshold of 3 photoelectrons occuring in 
coincidence with neutron events in the neutron counter were accepted. 
We developed a low noise circuit and system and were able to detect signals 
down to 3 photons under a vast amount of noise from the accelerator. 


The pulse heights from the CaF$_{2}$, BC501A, and NaI(Tl) detectors 
and the timing information of the BC501A, CaF$_{2}$, and the deuteron 
beam pulses as determined by an inductive pickup coil were 
recorded event-by-event through CAMAC-SFVME-SUN(Spark Station) on 
a disc. 
The trigger rate was typically 
about 100 cps, which contained a few cps of true signals. 

\section{\bf Results}

Energy spectra for the CaF$_{2}$ detector were obtained in coincidence 
with the neutrons selected by the PSD method. 
In the WIMPs search, events from the central CaF$_2$(Eu) crystal are selected 
by setting a proper window on the roll-off ratio spectrum \cite{wepa}, 
which gives substantial reduction of backgrounds. 
Roll-off ratio is defined by the relative pulse height difference 
between the left and right PMT to identify the firing position in the 
CaF$_2$ detector. 
The light guide of CaF$_2$(pure) crystals acts 
not only as a light guide but also as an active veto counter for the 
central Eu doped one. 
We applied the same cut used for the roll-off ratio. 

An energy spectrum for CaF$_{2}$(Eu) 
at a neutron scattering angle of 60$^{\circ}$ is shown in Fig.~\ref{spect}(B). 
The two peaks correspond to the Ca and F recoils with energy of 91 keV and of 192 keV, respectively. 
The energy spectra were also obtained at 30$^{\circ}$, 40$^{\circ}$, and 50$^{\circ}$. 
The peaks were fitted with a Gaussian 
to determine the mean and the width of the observed energy distribution.
The widths of the Gaussians used in the fit include 
the measured energy resolutions of the detector, 
the uncertainty in the recoil energy which is caused by the 
spread in neutron beam energy, 
and the spread in the scattering angles. 
The relative intensities of the two peaks versus scattering angles 
of neutron are consistent with calculations based on the known elastic 
scattering cross sections. 
It is noted here that the energy threshold of CaF$_2$ detector is only 
a few keV, which is low enough to observe the low recoil energy. 

The fluorine peak has a tail on the high energy side 
in Fig.~\ref{spect}(B). 
The tail may be related to $\gamma$-rays from the 
recoiling nucleus that Compton scatter in 
the target scintillator depositing energy in addition to the recoil energy. 
Since the CaF$_2$ is a low $Z$ (atomic number) material, the photoelectric 
effect is of minor importance. 
$^{19}$F has two low-lying excited levels at 109.9 and 197.2 keV. 
The differential cross sections for inelastic scattering for these excited 
states are substantial according to the NNDC database \cite{nndc}. 
A simulation with GEANT shows that 21 $\%$ of $\gamma$-rays 
originating from these excited states interact within the crystal 
through Compton scattering. 
We observed that this tail was reduced by several $\%$ by requiring 
anticoincidence between the upper and lower NaI(Tl) counters. 
This is consistent with an estimate, which includes NaI's acceptance and cross 
sections at each angle. 

Fig.~\ref{f}(A) shows the measured energy in keV for Ca and F recoils.  
The obtained f-values are also shown in Fig.~\ref{f}(B) and (C) together with the previous results of the UK Collaboration \cite{UK98} and the BPRS Collaboration \cite{BPRS}. 
The dominant contributions to the vertical error bars of the present result, 
Fig.~\ref{f}(B) and (C), are 
due to the CaF$_2$(Eu) detector resolution and the uncertainty in the 
calculated recoil energy, which originates from the beam energy spread 
due to energy loss at the primary target. 

\section{\bf Discussion}

Our result indicates that the f-value of CaF$_2$(Eu) crystal 
increases with decreasing recoil energy, 
which is consistent with the result of the UK Collaboration \cite{UK}\cite{UK98}. 
On the other hand, the BPRS Collaboration 
\cite{BPRS} suggests that the f-value does not depend on the energy. 
The large f-value at low energy 
gives us a good sensitivity as a dark matter experiment.  
A low recoil energy is beneficial for a dark matter search. 
We applied two assumptions for the dependence of the f-value. 
Generally the f-value is large for particles with small $dE/dX$ (electrons) \cite{hion}. 
We then assume the f-value depends inversely on stopping power, which is represented as: 
\begin{equation}
f \propto \frac {1} {dE/dX} 
\propto \frac{1} {\frac{(dE/dX)_{F(Ca)}}{(dE/dX)_{\beta(\gamma)}}} 
\propto \frac{1}{\frac{(dE/dX)_{F(Ca)}}{\rm minimum-ionization}}. 
\end{equation}
Here $(dE/dX)_{F(Ca)}$ represents the stopping power of $^{19}$F ($^{40}$Ca), 
for which numerical values are tabulated by L. C. Northcliffe 
and R. F. Schilling \cite{ns}. 
In Fig.~3(B) and (C), the f-values calculated by eq. (2) 
are compared with the present experimental data. 
The above assumption gives good agreement with the observed data. 
For the dark matter search we extrapolated the f-value to the low 
energy region as shown in Fig.~3(B) and (C). 
Typically $dE/dX$ starts decreasing below the Bragg-peak, 
which is the energy region of interest. 
We therefore think the extrapolation to the low energy region is justified. 
Also shown in Fig.~3(B) and (C), for comparison, is a prediction from the Lindhard 
model \cite{lindhard}. 
It gives a good agreement with ionization detectors such as Ge and Si \cite{ge-si}, 
though this is inconsistent with our results. 

Our f-value is a little larger than that of the other measurements. 
This may represent a dependence on the Eu activator concentration of 
the CaF$_2$(Eu) crystal. 
We compared our result with the previous experimental results for CaF$_2$(Eu) whose Eu doping content is known \cite{UK}\cite{UK98}. 
The observed f-values for the fluorine nucleus at a recoil energy of 100 keV 
are shown in Fig.~\ref{act}. 
There is no expression 
for the f-value in terms of the activator concentration, 
though a theoretical expression for the 
relative luminescence efficiency as a function of activator concentration 
is given \cite{jw}\cite{ej}\cite{mm}. 
If we assume that we can apply it to the f-value also, then we obtain 
the curve shown in Fig.~\ref{act}. 
Here the f-value is represented as, 
\begin{equation}
f \propto \frac {c(1-c)^z} {c+\sigma_r(1-c)} 
\end{equation}
where $c$ is the mole fraction of the activator Eu, $\sigma_r$ is the ratio of 
the capture cross section for the exciting energy of luminescent at a lattice 
trapping (non-activator) site to that at an activator site, and $z$ is the number of lattice positions surrounding a given activator.  
The parameters for the curve in Fig.~4 are 150 and $2\times10^{-4}$ 
for $z$ and $\sigma_r$, respectively. 
Our values for the parameters turns out to be similar to 
those obtained in relative luminescence efficiency measurement of NaI crystal, which 
are 60 and 1$\times10^{-4}$ for $z$ and $\sigma_r$, respectively \cite{mm}.
However, there are 3 parameters including normalization for 3 data points, 
so more data is necessary. 
This assumption indicates that the f-value of our crystal would have an activator concentration close to that giving the maximum f-value. 
Since there is no clear foundation for using eq. (3) for the f-value,  
further studies are needed. 

\section{\bf Conclusion}

In summary, we observed recoils of calcium and fluorine nuclei 
in the range 25 $\sim$ 91 keV and 53 $\sim$ 192 keV in a CaF$_{2}$(Eu) 
crystal, respectively.
The scintillation efficiency was found to be 11 $\sim 20 \%$, and 9 $\sim 
23 \%$ for F and Ca, respectively.
These data serve to calibrate a CaF$_{2}$(Eu) dark matter detector whose 
results are discussed in Ref. \cite{wepa}. 



\begin{center}
\normalsize{\bf ACKNOWLEDGMENTS}
\end{center}
\vspace{15pt}

The authors thank Prof. Y. Nagai, Prof. M. Igashira, Dr. T. Shima, 
the Pelletron Accelerator group of Tokyo Institute of Technology, and Dr. H. Noumi 
for the help during the run, and the Nagai group for their kind support at TIT. 
The authors also thank Prof. Ejiri and Dr. T. Inomata for valuable discussions 
and K. K. Schaffer for careful readings. 
The present work is supported by the Grant-in-Aid of Scientific Research 
and the Ministry of Education, Science and Culture, Japan. 
The work of R. H. and H. M. was supported by JSPS Research 
Fellowships for Young Scientists. 
R. H. would like to acknowledge the hospitality of Nuclear Physics Laboratory, 
University of Washington. 




\newpage

\begin{figure}[h]
\caption[]
{Schematic top-view of neutron scattering apparatus. 
The neutron collimator was made from a cylindrical paraffin, 
60 cm in length with a tapered hole along the beam axis. 
The hole was 1 cm in diameter near the primary Tid$_2$ target and 3 cm in diameter 
near the secondary target, the CaF$_{2}$(Eu) crystal.
Each scintillator was viewed by a photomultiplier tube (PMT). }
\label{setup}
\end{figure}
\begin{figure}[hbt]
\begin{minipage}[t]{7cm}
\end{minipage}
\begin{minipage}[t]{7cm}
\end{minipage}


\caption[]
{(A): 
The typical distribution of the total integrated charge over 100 ns (which depends on 
scintillation decay time) versus the initial integrated charge over 26 ns 
(which is proportional to the rise time) recorded by the ADC. 
(B): 
An energy spectrum of CaF$_{2}$(Eu) 
at a neutron scattering angle of 60$^{\circ}$. 
The two peaks correspond to Ca and F recoils 
with energy of 91 keV and 192 keV, respectively.} 
\label{spect}
\end{figure}

\begin{figure}[hbt]
\begin{minipage}[t]{7cm}
\end{minipage}
\hspace{1cm}
\begin{minipage}[t]{7cm}
\end{minipage}

\caption[]
{(A): 
The measured energy in keV for Ca and F recoils in 
CaF$_{2}$(Eu) calibrated relative to electons ($\gamma$-rays) 
at each energy, versus the corresponding calculated recoil energy. 
The errors are statistical only. \\
(B) and (C): 
The f-values (quenching factors) 
versus recoil energy in CaF$_{2}$(Eu) for (B) Ca recoils and (C) F recoils. 
The crosses are the data from the present experiment, 
diamonds and stars are data points from the UK Collaboration \cite{UK98} 
and the BPRS Collaboration \cite{BPRS}, respectively.  
The f-values are not plotted by the BPRS Collaboration \cite{BPRS}, so 
we converted by taking only their vertical (statistical) error bar into account. 
The systematic errors are also included for the UK Collaboration and ours. 
The dashed line is the extrapolation of the f-values calculated by eq.~(2) 
and Ref. \cite{ns} (squares). 
The solid curve is the prediction from the Lindhard model \cite{lindhard}.} 
\label{f}
\end{figure}

\begin{figure}[hbt]
\hspace{2cm}
\caption[]
{The f-values as a function of Eu concentration (mole fraction) for fluorine nucleus  
in CaF$_2$(Eu). 
Experimental data from Ref. \cite{UK}(open circles) and this work(full 
circle) are shown, and the errors are only statistical. 
The dashed line is obtained from eq.~(3). }
\label{act}
\end{figure}

\newpage
\begin{figure}[hbt]
\vspace{14cm}
\hspace{-1cm}
\epsfxsize=18cm \epsfysize=10cm \epsfbox{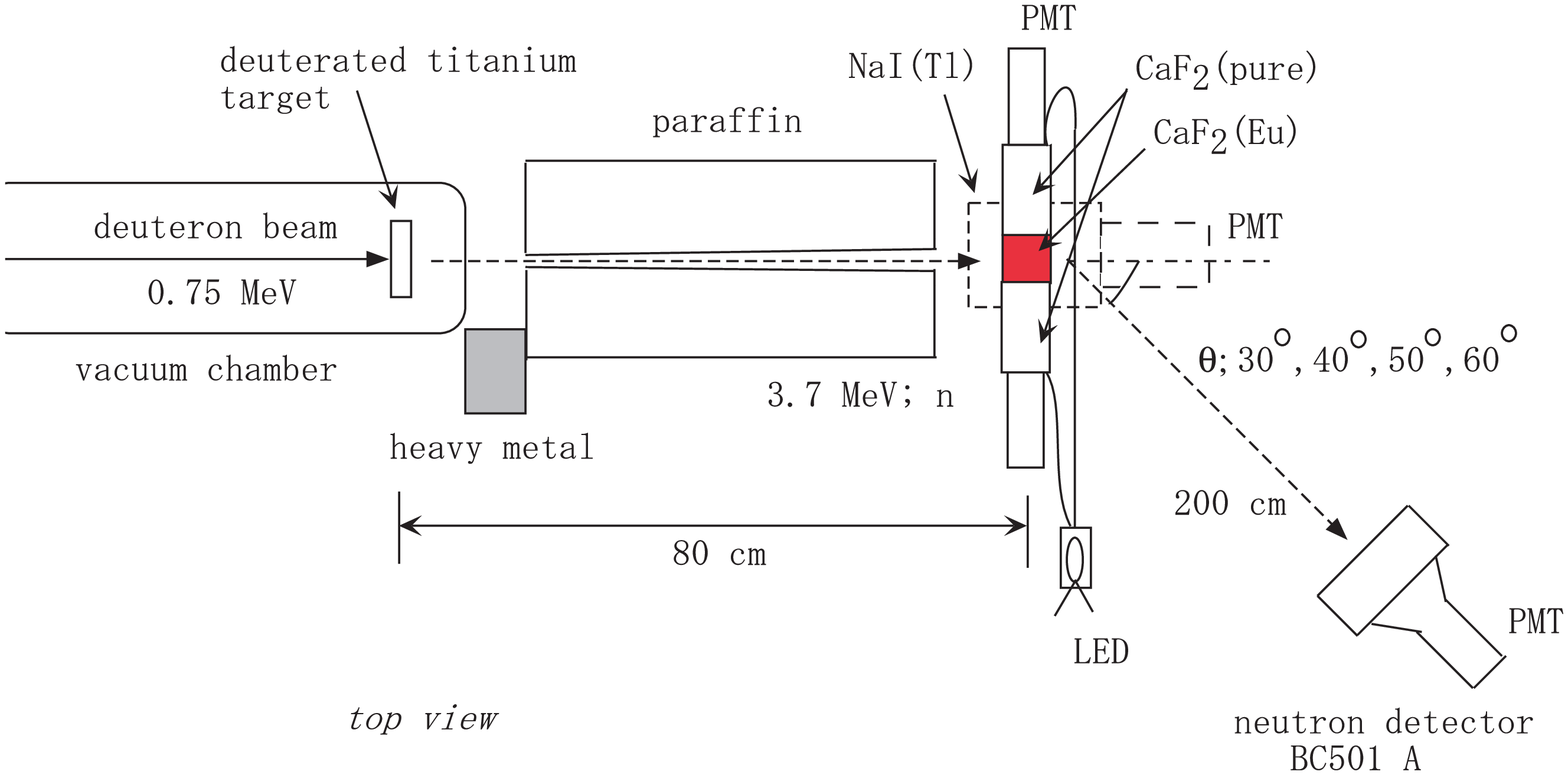}
\end{figure}
\vspace*{1cm}
{\bf Fig. 1}

\newpage
\begin{figure}[hbt]
\vspace{18cm}
\epsfxsize=15cm \epsfysize=15cm \epsfbox{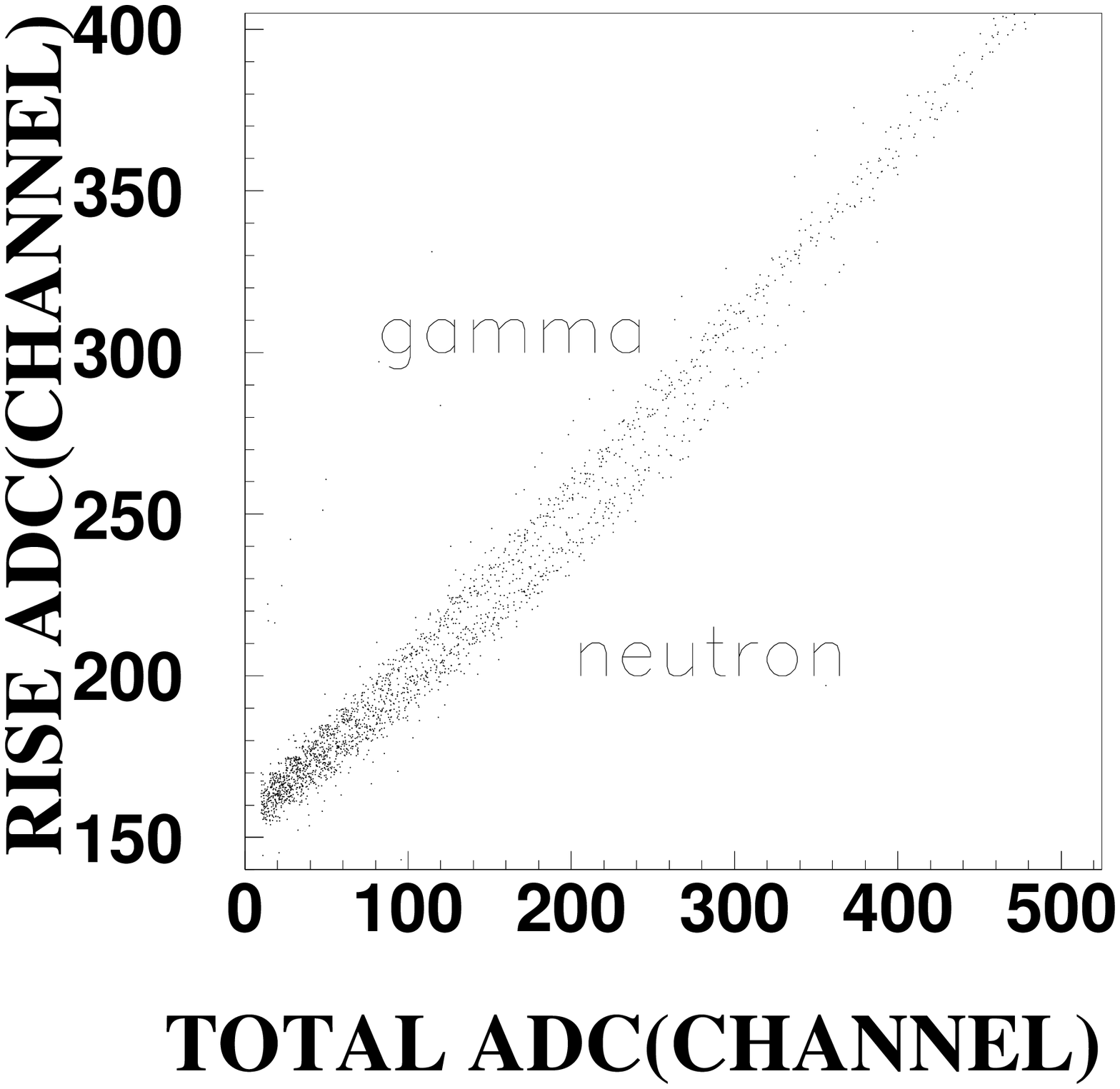}
\vspace*{1cm}
{\bf Fig. 2(A)}
\end{figure}

\newpage
\begin{figure}[hbt]
\vspace{18cm}
\epsfxsize=15cm \epsfysize=15cm \epsfbox{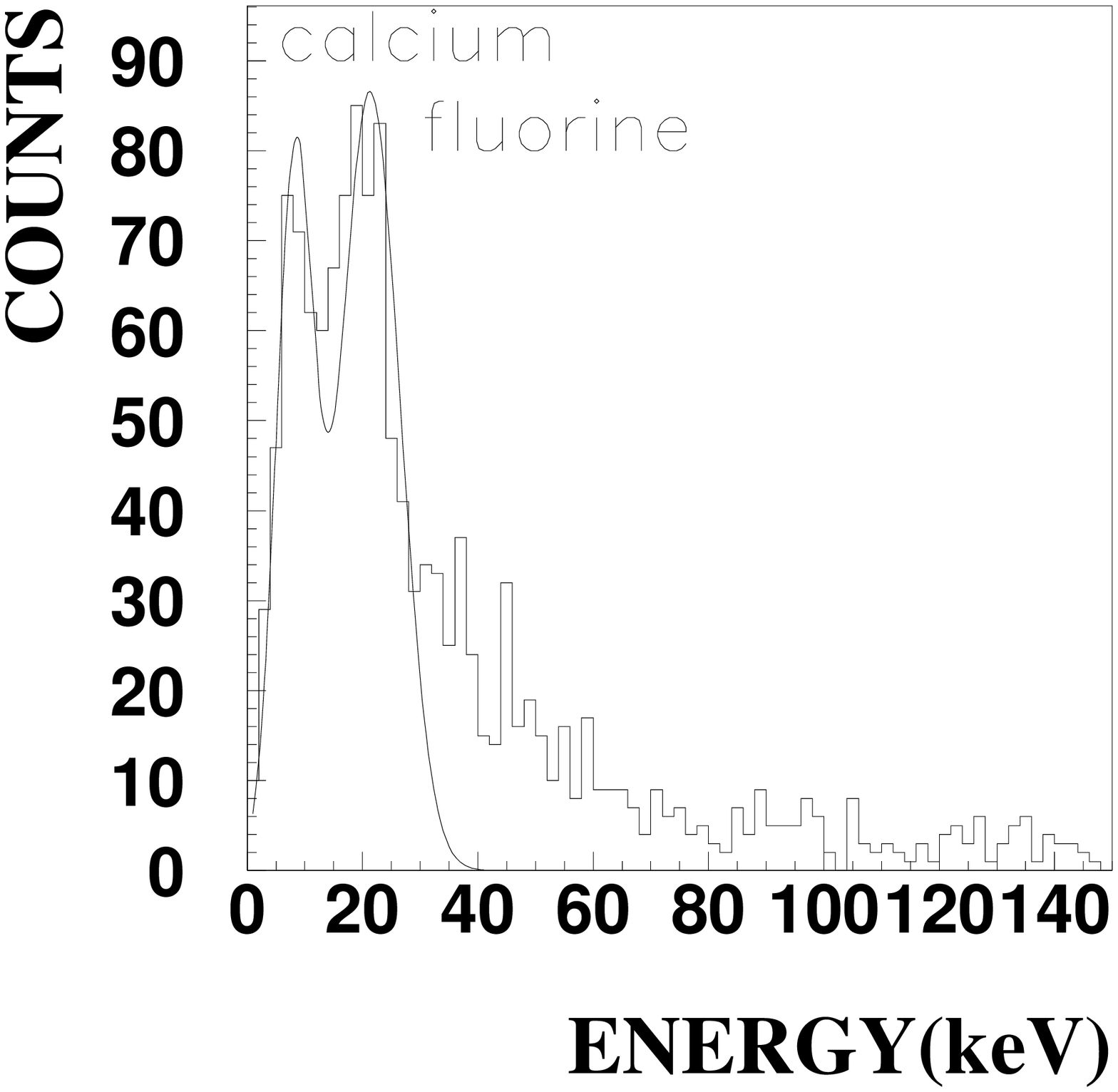}
\vspace*{1cm}
{\bf Fig. 2(B)}
\end{figure}

\newpage
\begin{figure}[hbt]
\vspace{18cm}
\epsfxsize=17cm \epsfysize=14cm \epsfbox{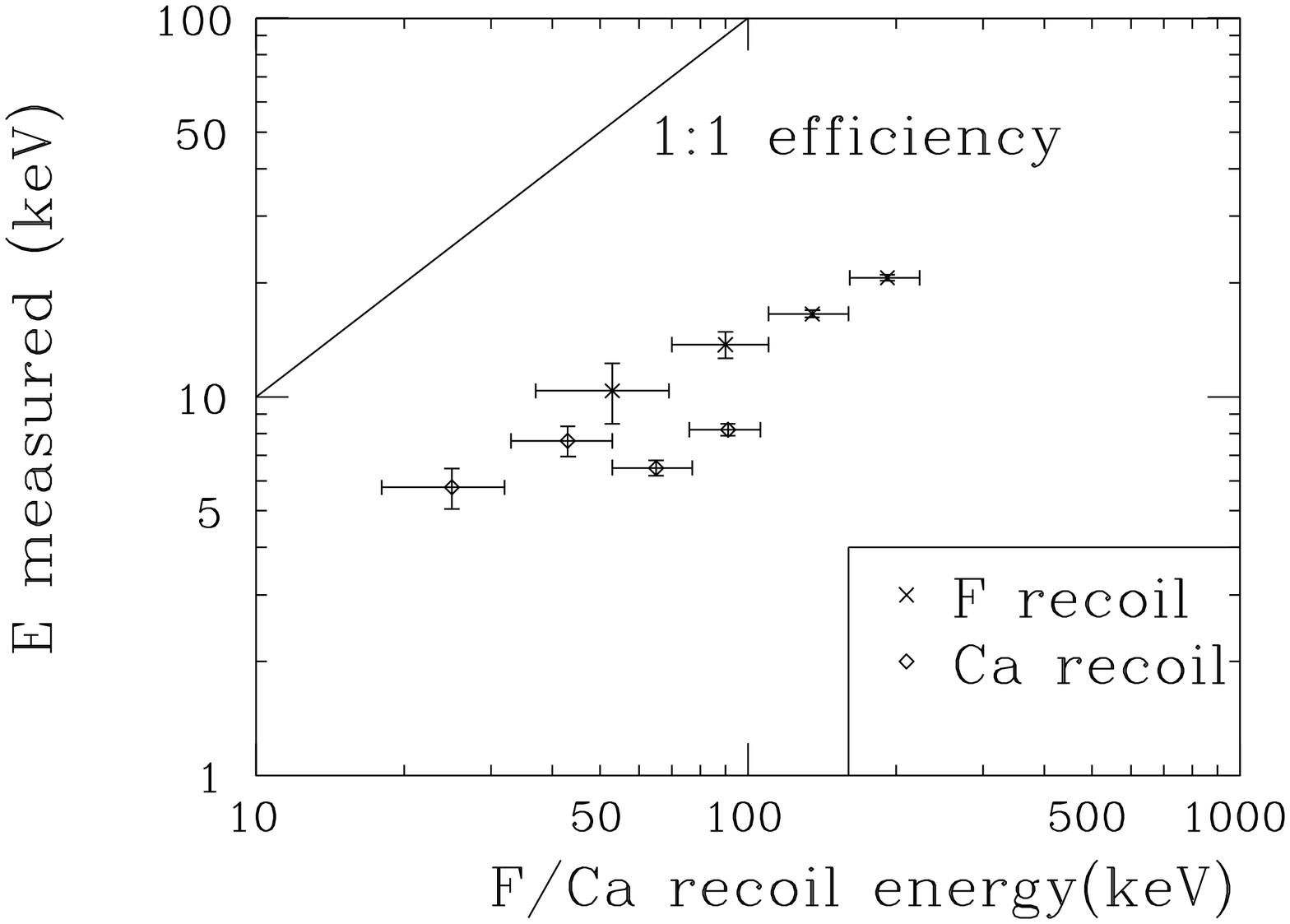}
\end{figure}
\vspace*{1cm}
{\bf Fig. 3(A)}

\newpage
\begin{figure}[hbt]
\vspace{18cm}
\epsfxsize=17cm \epsfysize=14cm \epsfbox{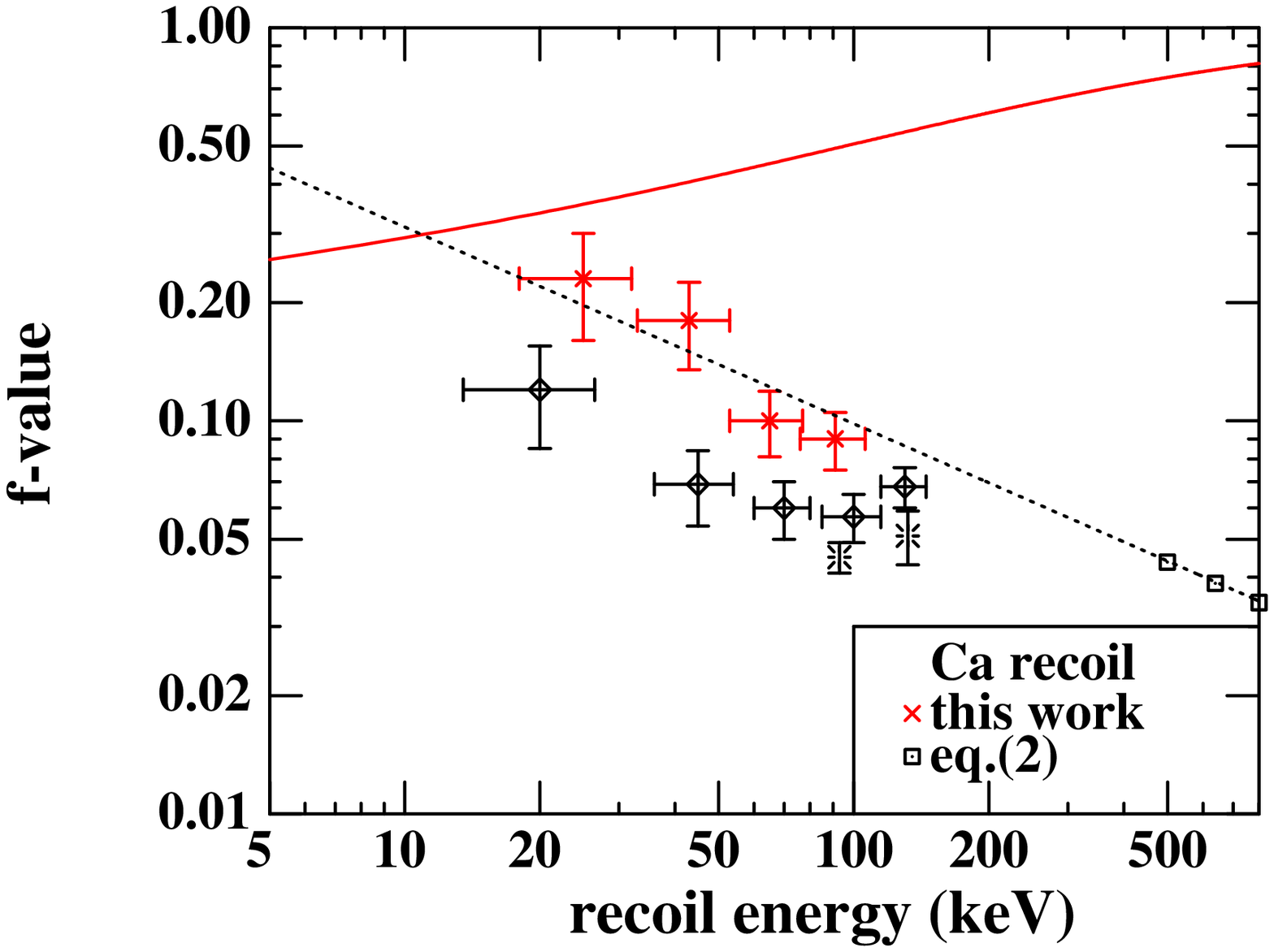}
\end{figure}
\vspace*{1cm}
{\bf Fig. 3(B)}

\newpage
\begin{figure}[hbt]
\vspace{18cm}
\epsfxsize=17cm \epsfysize=14cm \epsfbox{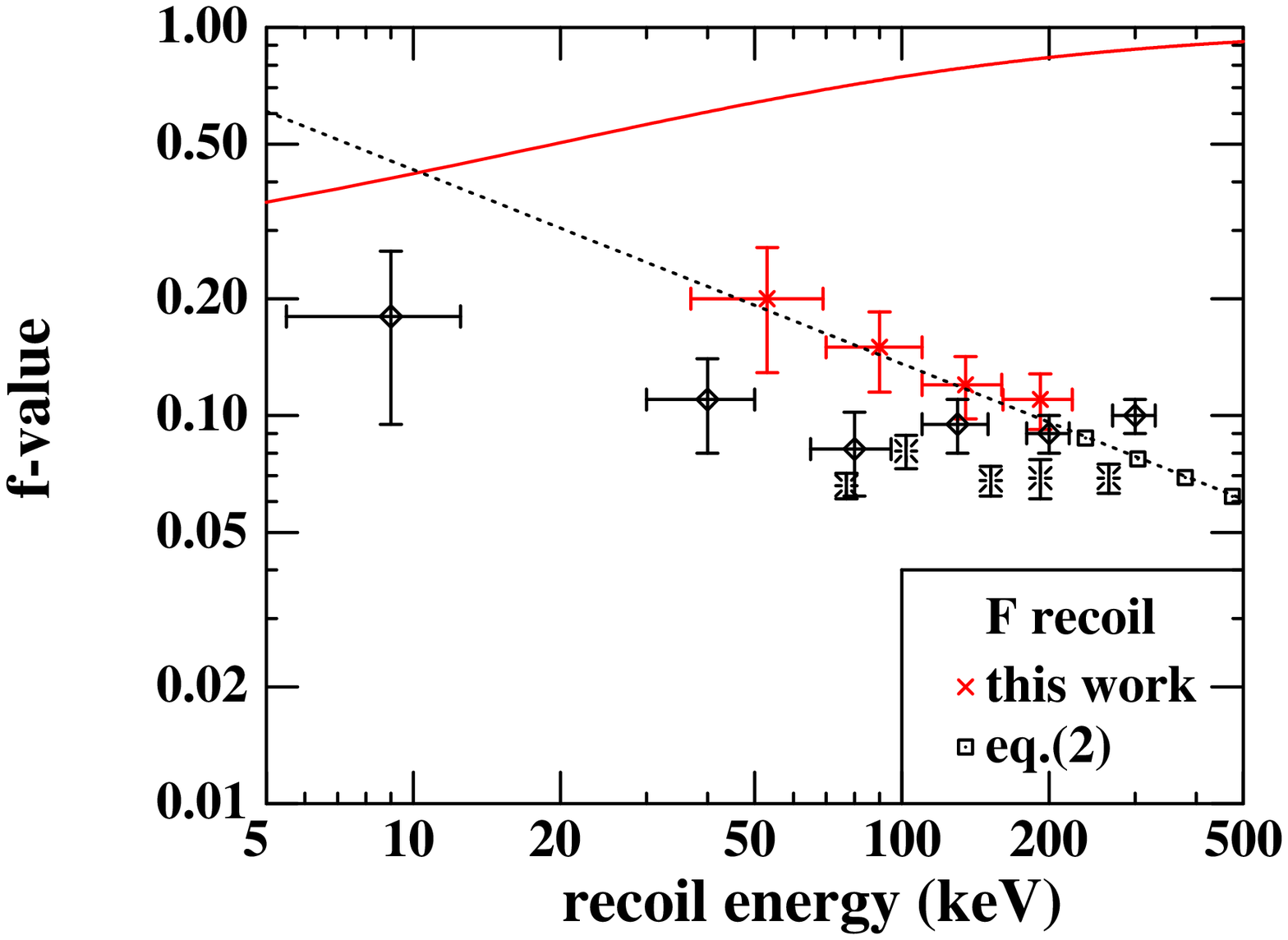}
\end{figure}
\vspace*{1cm}
{\bf Fig. 3(C)}


\newpage
\begin{figure}[hbt]
\vspace{18cm}
\epsfxsize=15cm \epsfysize=15cm \epsfbox{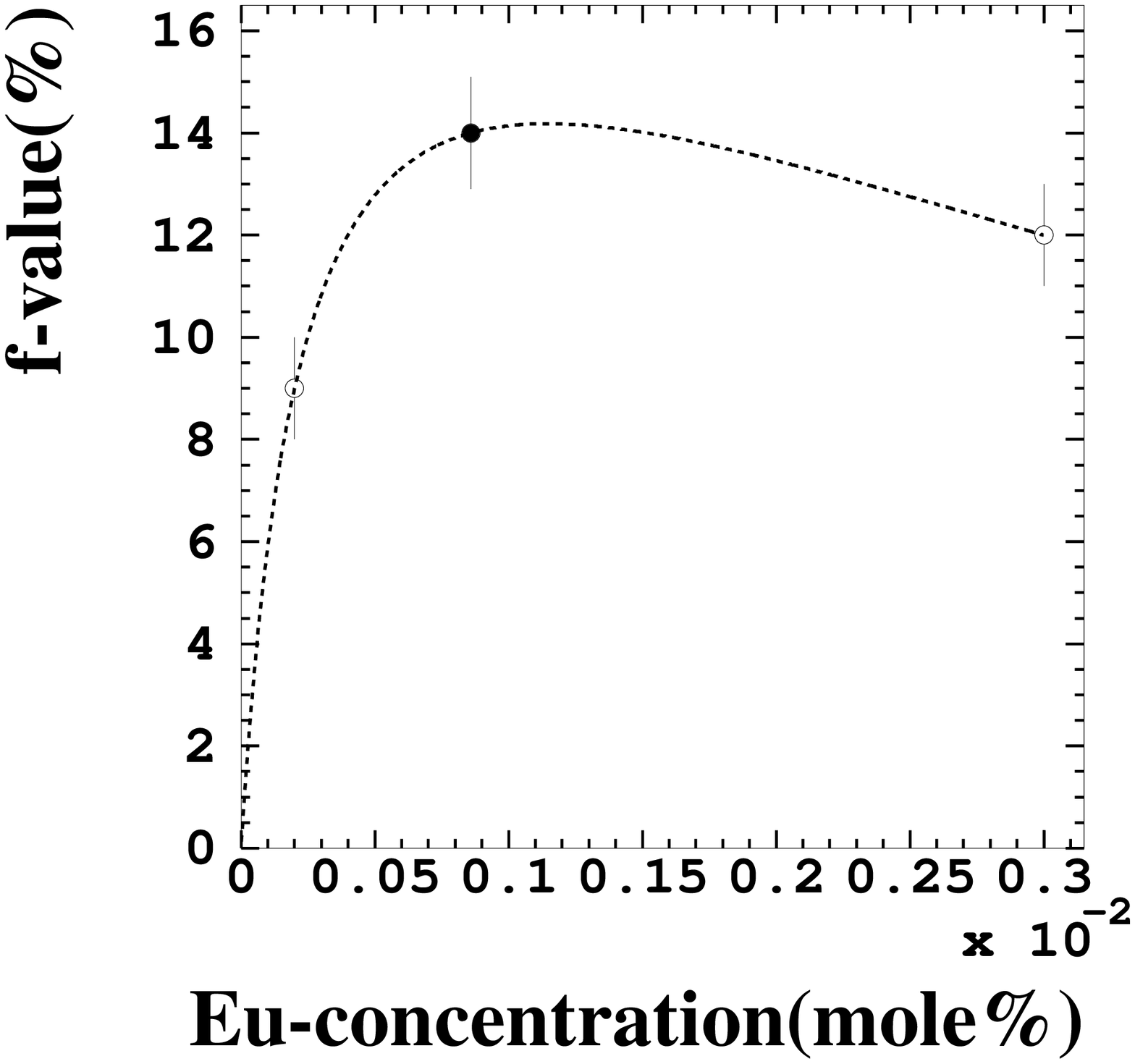}
\vspace*{1cm}
{\bf Fig. 4}
\end{figure}

\end{document}